\documentclass{llncs}
\usepackage[T1]{fontenc}
\usepackage[utf8]{inputenc}
\usepackage{lmodern}
\usepackage[hyperref=auto,style=alphabetic]{biblatex}

\usepackage{amssymb}
\usepackage{fnpct}

\bibliography{nnexus} 
\bibliography{kwarc}
\usepackage{xspace}
\usepackage{paralist}
\usepackage{graphicx}
\usepackage{url}
\title{NNexus Reloaded}
\author{Deyan Ginev\inst{1} \and Joseph Corneli\inst{2}}
\institute{Computer Science, Jacobs University Bremen, Germany \and Knowledge Media Institute, The Open University, UK}
\date{\today}

\begin{document}
\maketitle
\begin{abstract}

Interlinking knowledge is one of the cornerstones of online
collaboration.  While wiki systems typically rely on links
supplied by authors, in the early 2000s the mathematics encyclopedia at
PlanetMath.org introduced a feature that provides automatic
linking for previously defined concepts.  The NNexus software suite
was developed to support the necessary subtasks of concept indexing,
concept discovery and link-annotation.
In this paper, we describe our recent reimplementation and revisioning of the NNexus system. 
\end{abstract}

\section{NNexus 1.0 -- Introduction}

PlanetMath.org is a mathematics digital library, built ``the commons-based peer production way'' \cite{krowne+building+2003}.  Like Wikipedia, which launched the same year, PlanetMath has been created by volunteer contributors from around the world.  However, unlike Wikipedia, PlanetMath focuses solely on mathematics.  Since its launch, it has used custom software both to support the display of mathematical expressions, and to facilitate the integration of new user-contributed content.  One of the features designed to assist in content integration was an \emph{autolinking service}.
This service allowed authors to write without
concerning themselves with wiki-style links to
technical concepts that had already been added to the corpus.
Instead, these links would
be added automatically -- and links would be
recalculated and adjusted automatically as the encyclopedia
grew, using a sophisticated caching and expiry system.
%
%
The system provided an example of named entity recognition \cite{nadeau2007survey},
where the entities to identify in submitted text are article titles,
the names of terms defined in the articles, and any known synonyms.
The process of adding links to named entities in text has come to be
known as ``wikification'' \cite{RRDA11}.

%
In 2006, NNexus 1.0 began the process of decoupling autolinking from
PlanetMath, and provided integration with other
corpora (Wikipedia, Mathworld) on a demonstration
basis \cite{gardner+nnexus:+2006}, an effort that has matured with the current release. 


\section{NNexus 2.0 -- Reload, Refresh, Refactor}
The primary goal of our rebuild was to decouple fully from the old
Noosphere system on PlanetMath.org.  A strong contributing motivation
 was that Noosphere was in the process of being deprecated on
PlanetMath and replaced by the new Planetary system \cite{CorneliDumitru:OpenMathUIWiP2012}.  The new NNexus works with Planetary, but also functions in a stand-alone fashion, and is published as a software library on the Comprehensive Perl Archive Network (CPAN)\footnote{Run \textbf{cpan NNexus} to install the software locally.}.  It has been refactored to operate either as a web service\footnote{A demonstration instance is available at \url{http://nnexus.mathweb.org}.\\ Sending an HTML snippet as the body of a POST request will return the link-annotated snippet back, embedded in a thin JSON wrapper.}, or programmatically via an API.
NNexus accepts arbitrary HTML input and performs concept
discovery against its concept index, followed by a
serialization of the mined data, either as stand-off metadata or by in-place embedding.




%


Concept indexing is performed by NNexus' built-in web crawler. It is based on a plugin
architecture.  Every indexed web resource requires its own
indexer class, which contains the custom rules for
detecting the concept definitions in the page.  For
example, PlanetMath's key terms are found in RDFa metadata that has
been deposited in the encyclopedia pages, whereas the
Digital Library of Mathematical Functions lists its
defined concepts in its index as bold-anchored elements.

The current NNexus release ships with a database that
integrates the concepts from seven web resources for
mathematical concepts.  These include the three best-known
web resources for mathematics -- Wolfram's MathWorld; 
PlanetMath.org; and Wikipedia -- as well as Springer's
Encyclopedia of Mathematics; the Digital Library of Mathematical Functions
(DLMF); the nLab (which focuses on category theory); and
the recently created MathHub.info.

 At the time of writing, the NNexus index contains just under
50,000 unique concepts in its index.  With the introduction of
client-side tools for embedding NNexus \cite{Ginev:CICM-WS-WiP2013},
we can also report successful
auto-linking in third-party platforms such as arXiv.org and Zentralblatt
MATH. 
\section{Concept Discovery}
The NNexus implementations to date have only scratched the surface of the
knowledge discovery problem.  NNexus performs longest-token matching,
aided by classic preprocessing techniques (stopword lists,
morphological normalization) to discover all possible concept
candidates. Concepts are considered discovered if there is an exact match between the linguistically normalized input document and the identically normalized concept index. When a concept $A$ is a substring of a concept $B$, since they both match at the same starting point of the input, preference is given to the longer string $B$. To demonstrate, take $A$ to be ``fundamental groupoid'' and $B$ to be ``fundamental groupoid functor''.

This simplistic approach leads to false positive hits, for instance,
in words that have multiple part-of-speech uses or words that have both technical and
everyday meanings.  Accordingly, each of the following examples becomes a candidate for linking, even though the words in the right-hand column are not being used in a technical sense.
\begin{center}
\begin{tabular}{ll}
``\textit{Let $G$ be a \textbf{group}}'' & ``\textit{\textbf{group} the numbers in rows}'' \\
``\textit{\textbf{chain} in a graph}'' & ``\textit{\textbf{chain} made of steel}'' \\
``\textit{\textbf{permanent} of a matrix}'' & ``\textit{using a \textbf{permanent} marker}''  \\
\end{tabular}
\end{center}
This particular phenomenon is less observable as the length of the concept grows, both because longer words are less frequently overloaded and because multi-word concepts rarely have non-technical meanings. Inversely, the problem is particularly challenging in concepts with short single-word names.

Another challenge is disambiguating between overloaded concept names, used differently in different scientific areas. To address that, NNexus does not immediately return all of the named
entities it discovers.  Instead, it first uses a clustering algorithm, based on a distance metric between the classes of the MSC \cite{AMS:MSC2010} categorization scheme, which determines
a kernel of closely related concepts. We have observed our distance metric is effective in separating concepts from typically disjoint subfields of science, but less successful in making fine-grained distinctions in subfields that tend to have a lot of mutual connections. For example, ``entanglement'' is a concept both in quantum mechanics and graph theory.  NNexus is able to tell which meaning is intended by contextually clustering with the rest of the discovered concepts.  However, generic concepts such as ``equivalence'' tend to be redefined in closely related subfields, and NNexus cannot tell these apart.
%

In addition to these technical limitations, NNexus is limited by the quality of the metadata provided by
its indexed sources.  For example, as pointed out by one of the
reviewers of this paper, PlanetMath currently has no
article on ``classical logic'', and links to this term are
currently being directed to PlanetMath's article on
quantum logic.  This looked like a rather strange error
until we realized that the quantum logic article includes
a definition of the term ``classical logic'', in
contravention of the ``one main concept per article''
norm.

\section{NNexus 3.0 Revolution -- an Outlook}

Auto-linking continues to be a useful tool around
PlanetMath.  For instance, Planetary added support for
contributing problems and problem sets, and technical
terms in problems are linked to definitions drawn from the
encyclopedia.  We plan to add a PlanetMath feature where, given a
contributed piece of text, a small ``course packet'' of
preliminaries would be built on the fly, created out of
auto-linked encyclopedia articles.  Thanks to the metadata in
the links provided by NNexus, we will be able to consider
both ``incoming'' and ``outgoing'' links -- this means we
can discover applications of a concept as well as simpler concepts.
%

Some other efforts that we plan to explore include
autolinking in math blogs, such as the blogs indexed
on \url{http://mathblogging.org/}.  NNexus could build a ``term cloud'' of technical terms from across the
math blogosphere, providing a useful access method that
parallels the familiar tag cloud. 

The main challenge ahead is to solve the problem of
reliable concept discovery. The immediate goal is to
achieve reliable disambiguation of overloaded concept
words (such as ``set'' or ``group''), possibly by employing the help of a part-of-speech tagger. A complementary idea is to improve precision by augmenting
longest-token matching with weights derived by statistical
term-likelihood analysis.  As statistical
term-likelihood methods do not depend on an a priori fixed
lexicon, they could also be used to detect concepts that are
not yet included in the index.  That would allow us to
enable another desirable feature -- the automatic creation
of dangling links (similar to Wikipedia's ``red links'').

Deeper scrutiny of mathematical formulas and terms will
allow us to link occurrences of math constants in MathML,
both globally, for symbols like the reduced Planck constant
$\hbar$, and locally, following the annotation of the
corresponding natural language term, such as ``Assume a
cyclic group $\mathbb{Z}_{mn}$ \ldots''.  

\printbibliography
\end{document}